%% file: ms.tex
\documentclass[sigplan,10pt]{acmart}
\AtBeginDocument{%
  \providecommand\BibTeX{{%
    \normalfont B\kern-0.5em{\scshape i\kern-0.25em b}\kern-0.8em\TeX}}}
    
    \setcopyright{acmcopyright}
\copyrightyear{2021}
\acmYear{2021}
\setcopyright{acmcopyright}\acmConference[EuroMLSys'21]{The 1st Workshop on Machine Learning and Systems}{April 26, 2021}{Online, United Kingdom}
\acmBooktitle{The 1st Workshop on Machine Learning and Systems (EuroMLSys'21), April 26, 2021, Online, United Kingdom}
\acmPrice{15.00}
\acmDOI{10.1145/3437984.3458830}
\acmISBN{978-1-4503-8298-4/21/04}

\usepackage[utf8]{inputenc} 
\usepackage[T1]{fontenc}    
\usepackage{hyperref}       
\usepackage{url}            
\usepackage{booktabs}       
\usepackage{amsfonts}       
\usepackage{nicefrac}       
\usepackage{microtype}      
\usepackage[]{graphicx}
\usepackage{color}
\usepackage{caption}
\usepackage{subcaption}

\definecolor{red3}{rgb}{0.80,0.00,0.00}

\begin{document}
\title{Towards a General Framework for ML-based Self-tuning Databases}

\author{Thomas Schmied$^{\ast}$, Diego Didona, Andreas D{\"o}ring, Thomas Parnell, Nikolas Ioannou}
\email{e1553816@student.tuwien.ac.at, {ddi,ado,tpa,nio}@zurich.ibm.com}
\affiliation{%
  \institution{IBM Research - Zurich}
  \country{Switzerland}
}
\renewcommand{\shortauthors}{Schmied, et al.}

\thanks{$^{\ast}$Now at Vienna University of Technology, Austria.}
\begin{abstract}
\input{abstract}

\end{abstract}


\maketitle

\input{introduction}
\input{background}

\input{challenges}

\input{evaluation}
\input{conclusion}
\bibliographystyle{ACM-Reference-Format}
\bibliography{ms}

\end{document}

%% file: abstract.tex

Machine learning (ML) methods have recently emerged as an effective way to perform automated parameter tuning of databases. 
State-of-the-art approaches include Bayesian optimization (BO) and reinforcement learning (RL).
In this work, we describe our experience when applying these methods to a database not yet studied in this context: FoundationDB. 
Firstly, we describe the challenges we faced, such as unknown valid ranges of configuration parameters and combinations of parameter values that result in invalid runs, and how we mitigated them.
While these issues are typically overlooked, we argue that they are a crucial barrier to the adoption of ML self-tuning techniques in databases, and thus deserve more attention from the research community.
Secondly, we present experimental results obtained when tuning FoundationDB using ML methods.
Unlike prior work in this domain, we also compare with the simplest of baselines: random search.
Our results show that, while BO and RL methods can improve the throughput of FoundationDB by up to 38\%, random search is a highly competitive baseline, finding a configuration that is only 4\% worse than the, vastly more complex, ML methods. 
We conclude that future work in this area may want to focus more on randomized, model-free optimization algorithms.

%% file: introduction.tex
\begin{figure}[b!]
\centering
\includegraphics[scale=0.4]{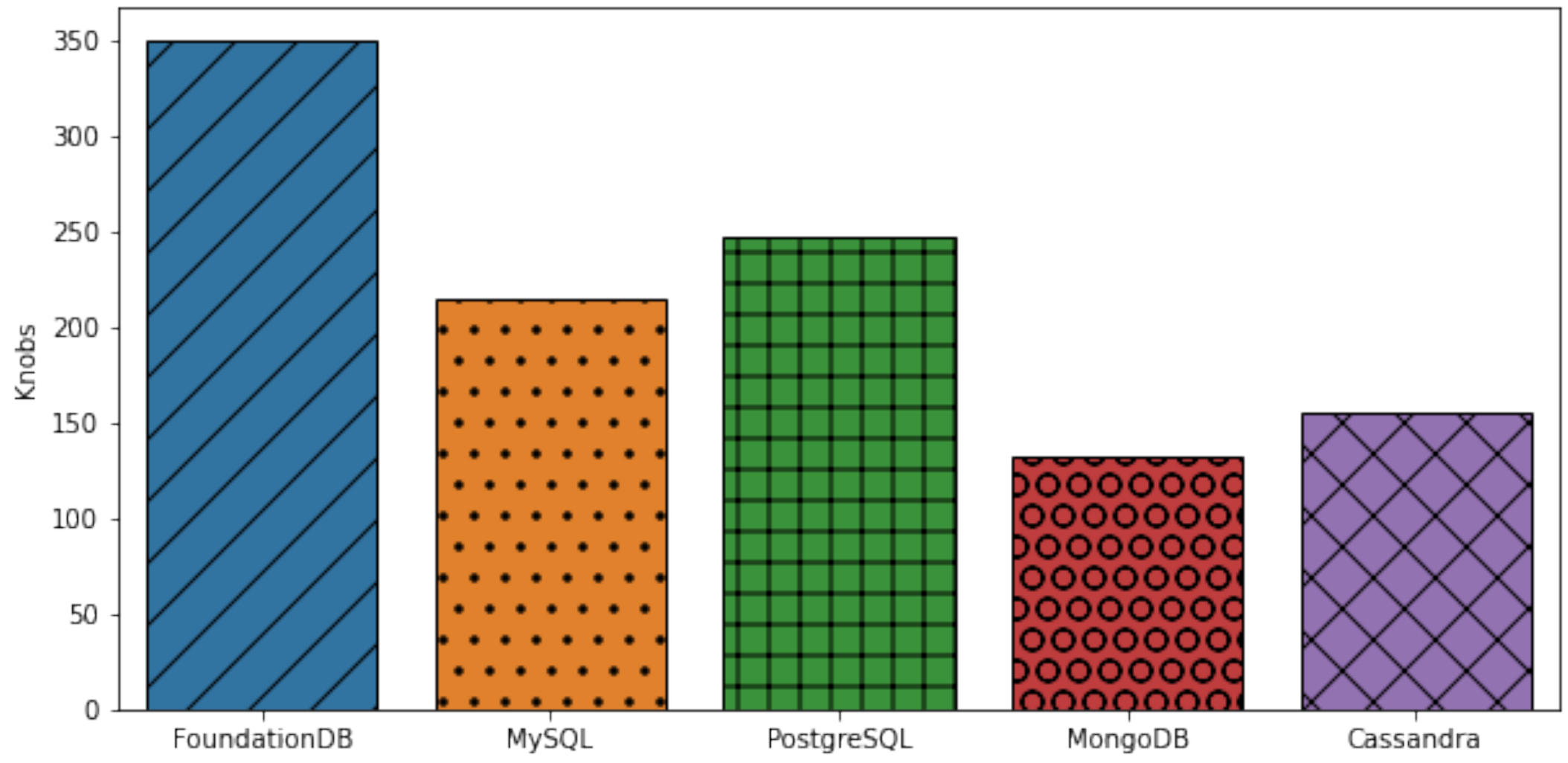}
\caption{Number of tuning parameters exposed by several established databases.}
\label{fig:intro}
\end{figure}
\section{Introduction} 
Optimizing the configuration space of database systems is crucial in improving the performance and latency  experienced by their users, as well as for reducing their operational costs.
However, optimizing the performance of a database system
is far from trivial, 
since modern databases expose tens to few hundreds of tuning parameters, resulting in a high-dimensional exploration space with significant performance variability across configuration points. Figure~\ref{fig:intro} reports the number of tuning parameters exposed by  established systems such as FoundationDB~\cite{fdb}, MySQL~\cite{mysql}, PostgreSQL~\cite{postgresql}, MongoDB~\cite{mongodb}, and Cassandra~\cite{cassandra}.
Over the last years, machine learning (ML) has emerged as a promising technique to automatically tune database configurations. Two state-of-the-art ML approaches to database tuning are Bayesian optimization (BO)~\cite{brochu2010tutorial} and reinforcement learning (RL)~\cite{sutton2018reinforcement}, implemented, among others, by the Ottertune~\cite{ottertune} and CDBTune~\cite{cdbtune} systems, respectively.
The ultimate goal of these systems is to optimize the performance of a database assuming as little intervention and domain knowledge as possible from the database administrator.

In this paper, we discuss the challenges we have faced, how we overcame them, and the results we have obtained, in applying these two ML methods to tune FoundationDB, a distributed transactional database system that backs several critical workloads and cloud data services, such as Apple's CloudKit~\cite{fdb:sigmod19}, IBM Cloudant~\cite{cloudant}, and Snowflake~\cite{snowflake}.
We first show that applying existing ML self-tuning techniques to a new database, assuming little domain knowledge, poses several challenges; challenges that are typically overlooked or briefly discussed in related work and that are not accounted for by existing self-tuner prototypes.
For example, we show that identifying valid ranges for the parameter values is a critical step in enabling the database tuning process, and that applying existing off-the-shelf techniques is not enough to perform feature selection in a robust and effective fashion. We also describe how we have overcome these challenges in order to effectively deploy these self-tuning methods on FoundationDB.
Then, we analyze the effectiveness of the two methods in terms of the performance achieved by FoundationDB in transactions per second and compare them with a random search baseline.
We show that BO and RL methods perform similarly, both finding a configuration that can improve the average throughput of FoundationDB by 38\% with respect to the default configuration.
However, we also show that even simple random search based approach can find a configuration that is only 4\% worse than the best configuration found by the, vastly more complex, ML methods.
This result echoes recent findings in neural architecture search, that show that random search is competitive with more complex hyper-parameter tuning approaches~\cite{li2020random}.

Our experience provides twofold insights. First, research on ML-based approaches to self-tuning databases should not only focus on improving the quality of the employed ML models, but also on making the whole self-tuning process more robust and seamlessly applicable to diverse domains. Second, random search should always be considered as a baseline when evaluating a new approach, and new randomized approaches should receive more attention in future research works.

%% file: background.tex

\section{Background}
\noindent{\bf FoundationDB.} FoundationDB is an open-source, transactional, strongly consistent, distributed key-value store, and is used as a back-end data-platform for multiple data services. FoundationDB exposes more than 350 tuning parameters, which regulate the behavior of the system at several levels, including networking, storage, and transaction processing.

\noindent{\bf Machine learning approaches.}
We now introduce two among the most prominent ML approaches to database tuning, which are implemented by the solutions we investigate in this paper.

\noindent{\em $\bullet$ Bayesian optimization.} BO aims to optimize a target function that is unknown in closed form, and that is expensive to evaluate \cite{brochu2010tutorial, frazier2018tutorial}.
BO fits a surrogate model of the function to observed points and picks the next point to evaluate based on an acquisition function.
In database tuning, the function to optimize is the database performance (e.g., throughput), and each function point corresponds to a database configuration.
Ottertune~\cite{ottertune} is a state-of-the-art database tuning system based on BO.
It uses Gaussian processes to fit the surrogate model and uses expected improvement (EI) as the acquisition function.
EI carefully balances exploration (i.e., trying points of high uncertainty in the surrogate model) and exploitation (i.e., refining the search around known good points).

\noindent{\em $\bullet$ Reinforcement learning.}
In RL an agent interacts with an environment with the goal of maximizing its reward.
At every time step $t$ the agent is in state $s_t$ within the environment, performs action $a_t$, and receives reward $r_t$.
In database tuning, the environment corresponds to the database system to tune; the state $s_t$ encodes the internal metrics of the database at step $t$ (e.g., free memory and resource utilization); an action corresponds to setting the parameters of the database to a specific set of values; the reward $r_t$ is expressed as the improvement of the performance measured at step $t$ over the performance of the database with the initial, default configuration.
CDBTune~\cite{cdbtune} is a state-of-the-art database tuning system based on RL.
CDBTune implements Deep Deterministic Policy Gradient (DDPG) \cite{lillicrap2015continuous}, a recent RL algorithm based on neural networks.

%% file: challenges.tex
\section{Challenges and solutions}
We now describe the challenges we faced in applying BO and RL to optimize FoundationDB, and how we mitigated them.
Our goal is to apply the two techniques out-of-the-box as much as possible and assuming as little knowledge as possible about the internals of FoundationDB, as it would be the case for a non-expert in databases or ML. Although our discussion focuses on applying BO and RL to FoundationDB, our findings and observations apply also to other systems and other ML-based approaches to database tuning.
\subsection{Unknown tuning parameters and valid value ranges}
\label{subsec:val_range}
The first step towards applying a ML-based technique to optimize a database is identifying the parameters to tune and their value ranges. This is hard to do without domain knowledge. In previous work~\cite{kanellis2020too,cdbtune,ottertune}, the tuning parameters and their value ranges are determined by exploiting knowledge about the internals of the target database.  FoundationDB exposes the tuning parameters and their default values in a few source files. From those files, we programmatically extract 350 tuning parameters and their default values.

Setting the value ranges for the parameters poses some challenges.
On the one hand, defining small value ranges around the default values reduces the possibility of exploring some parameters widely enough, possibly leaving performance gains on the table. On the other hand, large value ranges around the default values pose two further challenges. First, they increase the size of the configuration space to explore, and hence the complexity of the optimization problem. Second, they are prone to defining invalid or unstable database configurations, which makes it hard to even identify the configuration space for which the database is in an operational state. In FoundationDB unstable configurations happen, for example, when setting some timeout parameters to very high values, which can make the system non-responsive. This issue of invalid configurations is not limited to FoundationDB, but also affects other data stores. For example, the popular RocksDB key-value store~\cite{rocksdb} may crash if the maximum depth and width of its internal tree data structure are set to high values~\cite{rocksdb-tuning, vldb2021}.

\begin{figure}[t!]
\centering
\includegraphics[scale=0.5]{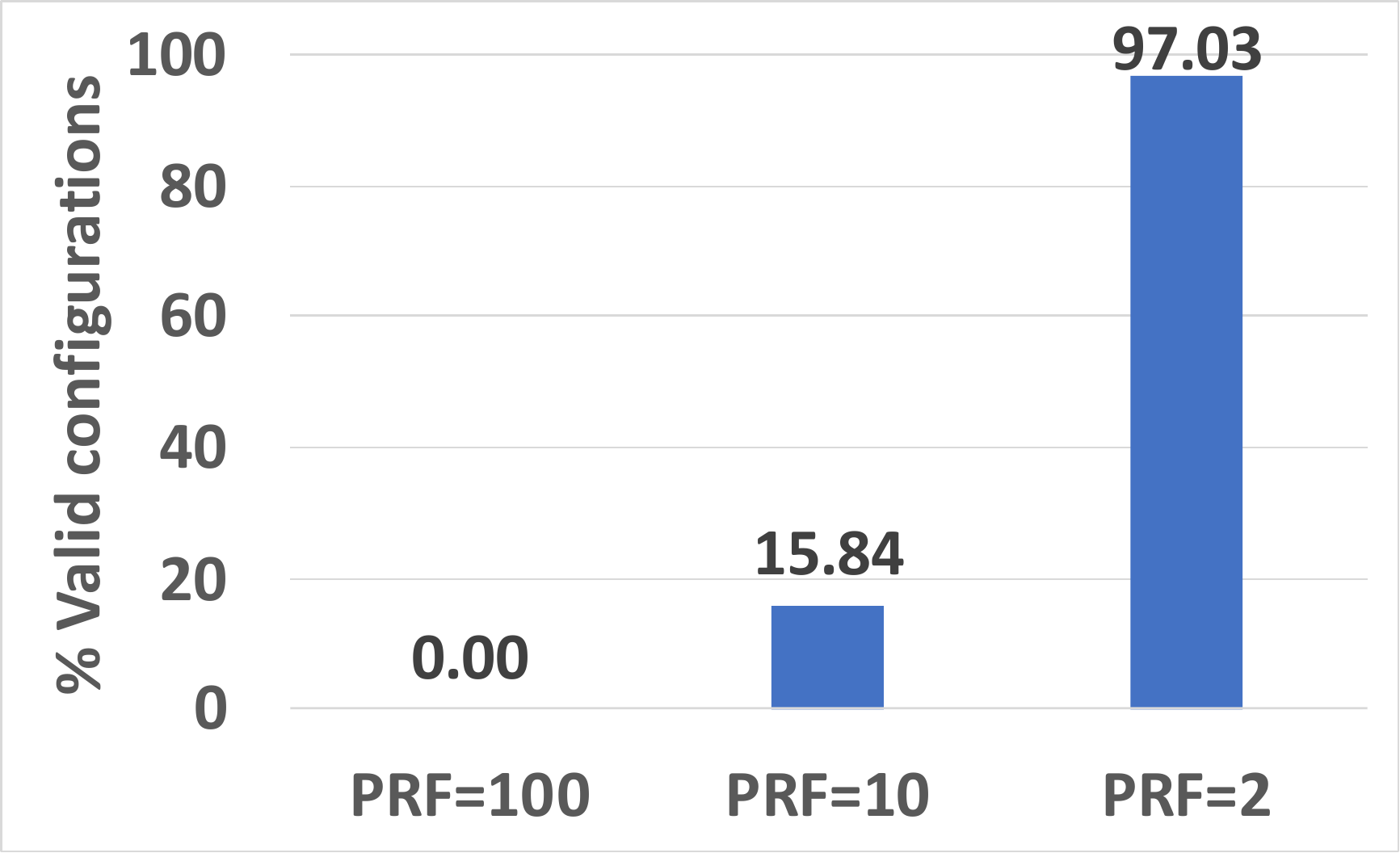}
\caption{Percentage of valid random configurations obtained running LHS with different parameter range factors. The random configurations are obtained by applying LHS on all the parameters.}
\label{fig:prf}
\end{figure}

We investigate the issue with invalid configurations through a specific experimental study. We define the domain of each parameter by expressing its lower bound, resp. upper, bound by dividing, resp., multiplying, its default value by a scalar that we call {\em parameter range factor ($PRF$)}. Namely, we define the domain of parameter $p$ with default value $d_p$ as  $[d_p/PRF, d_p \times PRF]$. We set $PRF = 2, 10, 100$ to define small, medium and large value ranges.
For each $PRF$ value we generate $N=100$ different FoundationDB configurations, using  Latin hypercube sampling (LHS)~\cite{lhs}, which we describe next. In LHS, the domain of each parameter is split in $N$ equi-sized subdomains. For each subdomain a representative value is picked, so that each parameter has a candidate set of $N$ values. The $N$  configurations are then composed by drawing the value for each parameter at random and {\em without} replacement from the parameter's candidate set. This method ensures that for each subdomain of each parameter there is one configuration that contains a value in that subdomain.

Figure~\ref{fig:prf} shows that the percentage of valid configurations sharply decreases as $PRF$ increases.  With $PRF=100$, none of the configurations picked by LHS corresponds to a valid FoundationDB deployment. $PRF=2$ corresponds to almost all configurations being valid but has the aforementioned problem of exploring a very limited range of parameter values.
For the remainder of our study, we use $PRF=10$, which is low enough to produce some valid configurations, and high enough to explore a sufficiently large range of parameter values.

\begin{figure}[t]
\centering
\includegraphics[scale=0.5]{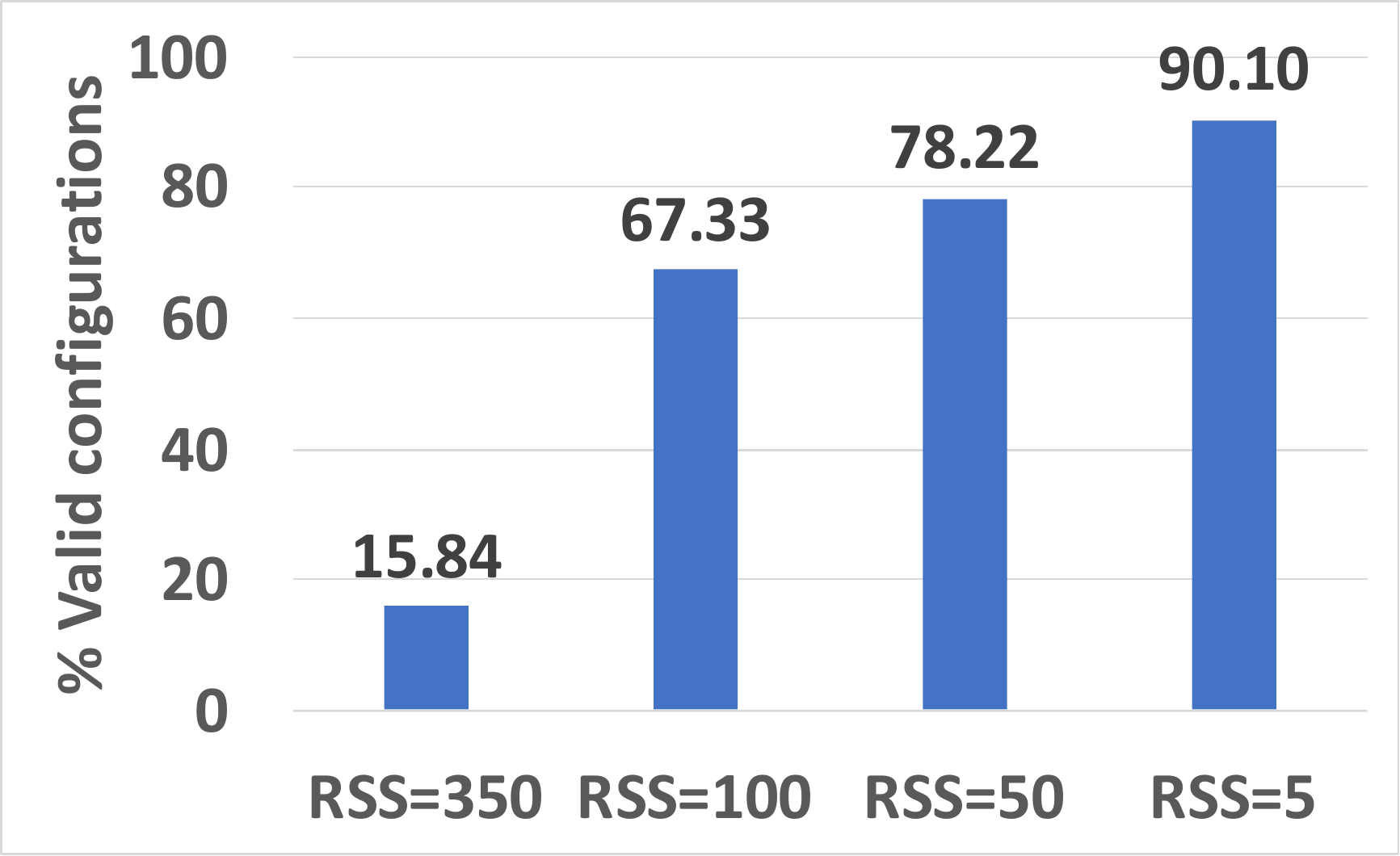}
\caption{Percentage of valid random configurations obtained running LHS with different random sub-sampling ($RSS$) values. The parameter factor range is set to 10.}
\label{fig:rss}
\end{figure}

\begin{figure*}[t!]
\begin{subfigure}[c]{0.45\textwidth}
\centering
 \includegraphics[scale=0.5]{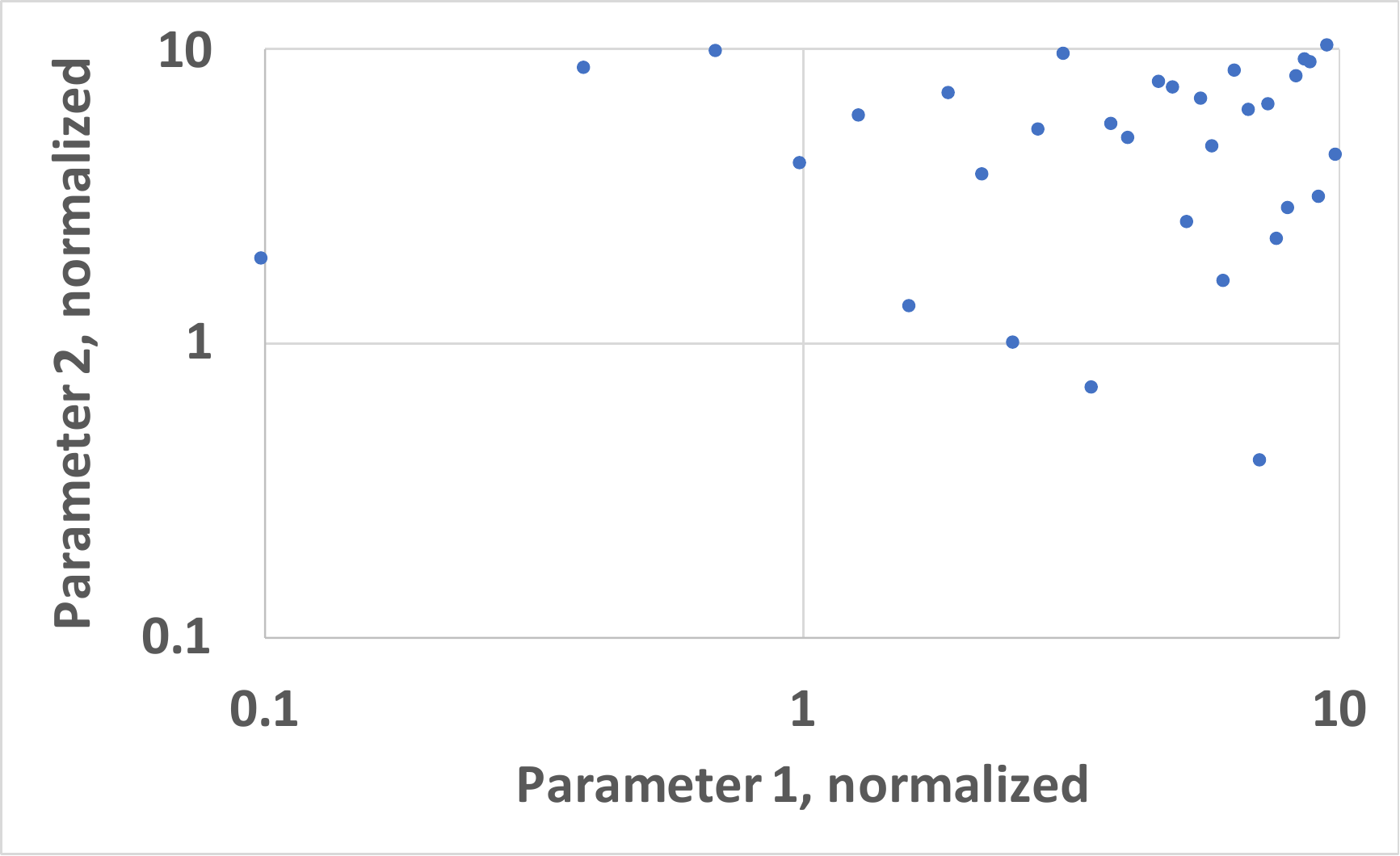}\caption{Standard LHS.}\label{fig:nonequi}
 \end{subfigure}
 \hfill
\begin{subfigure}[c]{0.49\textwidth}
\centering
 \includegraphics[scale=0.5]{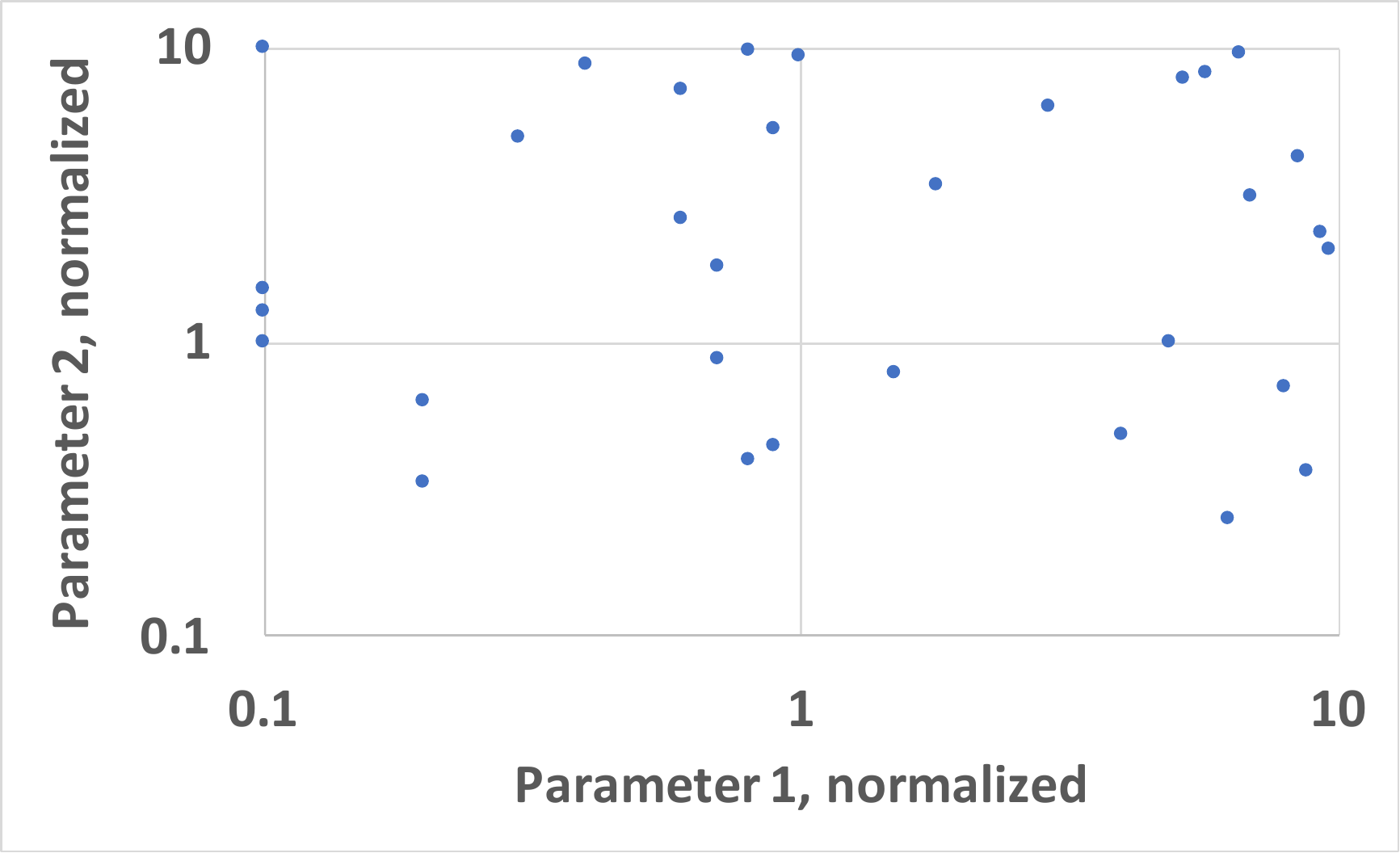}\caption{Symmetric LHS.}\label{fig:equi}
\end{subfigure}
\caption{Parameter space coverage of standard LHS (a) vs symmetric LHS (b). Axes are in logarithmic scale.  Symmetric LHS achieves a better sampling of a parameter's low values with respect to LHS, by sampling separately the [min,default) and [default,max] intervals.}
\label{fig:lhs}
\end{figure*}

\subsection{High-dimensional configuration space}
After defining the domain of the optimization problem, we aim to run BO and RL to maximize the performance of FoundationDB.  The optimization processes of BO and RL start by randomly sampling the parameter space to build an initial surrogate model of the performance function of FoundationDB.
However, as we have shown in the previous section, sampling across the whole configuration space of FoundationDB in a random fashion with a nontrivial $PRF$ results in the vast majority of configurations being invalid.
This hinders the effectiveness and the convergence speed of the optimization processes of BO and RL.
We note that it is arguably hard to pick randomly a valid configuration in a configuration space defined of over 350 parameters, with large value domains, with unknown semantics, and with highly intertwined performance and functionality dynamics.

To obtain a more reliable random sampling process while keeping a high enough value for $PRF$, we decide to reduce the size of the domain space by selecting a reduced set of parameters to tune -- a process known as feature selection in ML literature. Unfortunately, performing feature selection and identifying which are the parameters that are the most correlated with our optimization target, namely DB performance in transactions per second, again requires sampling the configuration space of FoundationDB, and suffers from the same issues described above.

~\\
\noindent{\bf Random sub-sampling.} To overcome this problem, we implement a variant of LHS that generates configurations where, for each configuration, only a random subset of the parameters has a non-default value. The random subset's size ($RSS$) is a parameter of this sampling approach.

\begin{figure*}
\centering
\includegraphics[width=1.0\textwidth]{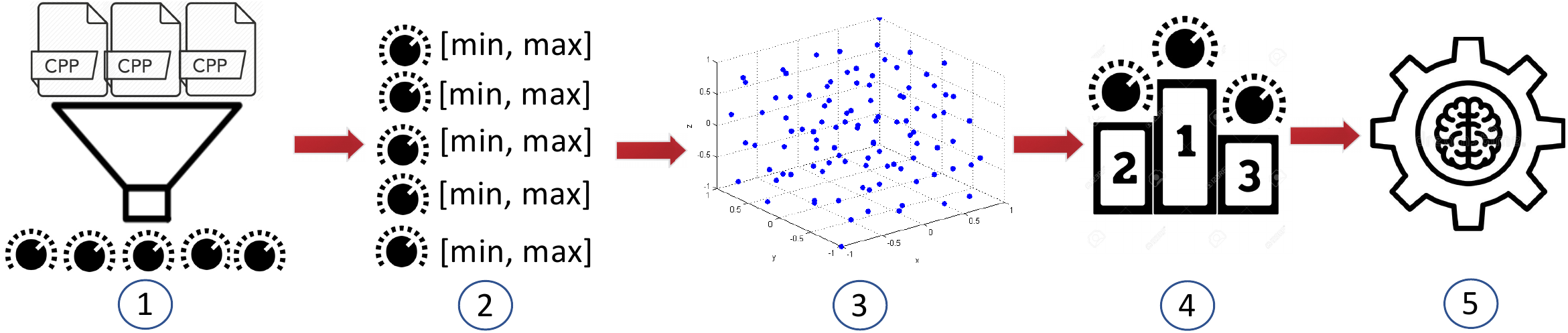}
\caption{Optimization pipeline. (1) Extraction of the tuning parameters from source code. (2) Definition of value ranges. (3) Random sampling of the configuration space. (4) Feature selection. (5) ML-based self-tuning.}
\label{fig:pipe}
\end{figure*}

Figure~\ref{fig:rss} reports the percentage of valid configurations obtained by our sampling approach with different $RSS$ values. As expected, lower values of $RSS$ yield a higher number of valid configurations, by employing fewer non-default values for the configuration parameters.
Based on these results, we choose to use $RSS = 50$ to perform feature selection. This value offers a good compromise between yielding a relatively high number of valid configurations and sampling from a large enough subset to capture joint performance dynamics of different parameters.

~\\
\noindent{\bf Symmetric LHS.}  To further improve the quality of the sampling process, we apply another modification to the LHS sampling approach. We note, in fact, that standard LHS splits the range of each parameter in equi-sized sub-ranges, and a single value is picked from each sub-range. When applied to the FoundationDB parameter value domains defined with multiplying and dividing a default value by $PRF$ (Sec~\ref{subsec:val_range}), LHS yields a skewed sampling of the parameter value ranges that is biased toward higher values. To illustrate the issue, assume that a parameter has a default value of 1, that  $PRF$ is 10, and the number of samples to draw at random is 100. Then, the parameter range is assigned a minimum value of 0.1 and a maximum value of 10.  Hence, LHS splits the [0.1, 10] range in 100 equi-sized sub-ranges. However, only 10 of these sub-ranges fall in the [mininum, default] range, and 90 fall in the (default,maximum] range.  This represents a problem in case variations around small values of the parameter yield significant performance variation.

To mitigate this issue, we implement a {\em symmetric} variant of LHS, that splits the [min, default] range and the (default, max] range in the same number of sub-ranges. Figure~\ref{fig:lhs} compares the configurations selected for two FoundationDB parameters by standard LHS (on the left) and our symmetric variant (on the right). To provide an easy visual comparison of the two approaches, we just report two parameters of a configuration,  the values of the parameters are normalized (so that the default value is 1), $PRF = 10$, and the axes are on logarithmic scale. As we can see, standard LHS draws far more configurations in the top right quadrant of the grid, that corresponds to configurations where both parameters have a value higher than the default one. Conversely, our symmetric LHS approach samples lower and higher values of the parameters more uniformly.

Our symmetric LHS is similar to logarithmic sampling, which samples the parameter range on a logarithmic scale.
Logarithmic sampling is used in ML  to tune the  hyper-parameters of an  algorithm  that span several orders of magnitude, e.g., the learning rate in neural networks.
Logarithmic sampling, however, by design draws more samples around the minimum of the parameter range in the linear domain, and  results in high values being under-represented. In the database tuning domain, this comes with the  risk of not capturing well enough the impact of high values of a given parameter. Symmetric LHS addresses this limitation  by sampling with the same frequency low and high values, and thus is more widely applicable to spaces defined over heterogeneous domains.

~\\
\noindent{\bf Feature selection.} We collect performance data corresponding to 200 configurations selected applying our variant of LHS that incorporates both the random sub-sampling and the symmetric sampling techniques. We use this data to perform feature selection using a random forest regressor and select the most impactful parameters according to the ranking output by the regressor. Similarly to recent work~\cite{kanellis2020too}, we find that a few parameters (10 in our case) account for the vast majority (90~\%) of the overall importance score. 
The parameters that the feature selection process identifies are the following: $\#$ of proxies, $\#$ of resolvers, \texttt{delay\_jitter\_offset},  \texttt{update\_storage\_process\_stats\_internal}, \texttt{slow\_smoo-}\\\texttt{thing\_amount}, \texttt{disk\_queue\_adapter\_max\_switch\_time},\\\texttt{shard\_bytes\_ratio}, \texttt{worker\_failure\_time}, and \texttt{txs\_pop-}\\\texttt{ped\_max\_delay}. 
We therefore use only these parameters to run the optimization phase of the two database tuning methods based on BO and RL.

%% file: evaluation.tex

\section{Experimental results}
We now describe the results we obtain when running the two self-tuning techniques. Figure~\ref{fig:pipe} depicts the full optimization pipeline we execute.

\noindent{\bf FoundationDB.} We use FoundationDB 6.2.20, using three server machines and two client machines. The machines are equipped with Intel(R) Xeon(R) CPU E5-2690 @ 2.90GHz CPUs, 251Gi of RAM, run Ubuntu 19.10 with a 5.3 Linux kernel, and are connected over a 100GBps network. We load the database with 1M key-value pairs. Each key is 16 bytes, and each value is 512 bytes. The clients generate transactions composed of 5 operations, according to a 60:40 read:write mix, and select keys according to a uniform random distribution.

\noindent{\bf ML techniques.} We use CDBTune to evaluate the RL-based approach, by adapting the code available at~\cite{cdbtune-git}. To evaluate the BO-based approach we use the BO implementation of scikit-optimize~\cite{scikit-optimize}. We configure the BO algorithm to use Gaussian processes, as in Ottertune. Last, we evaluate a random search approach based on scikit-optimize.
We let the three techniques run for 200 steps; we repeat each optimization process three times. Running the experiments took a total of 9 days.

\begin{figure*}[t!]
\begin{subfigure}[c]{0.45\textwidth}
\centering
 \includegraphics[scale=0.25]{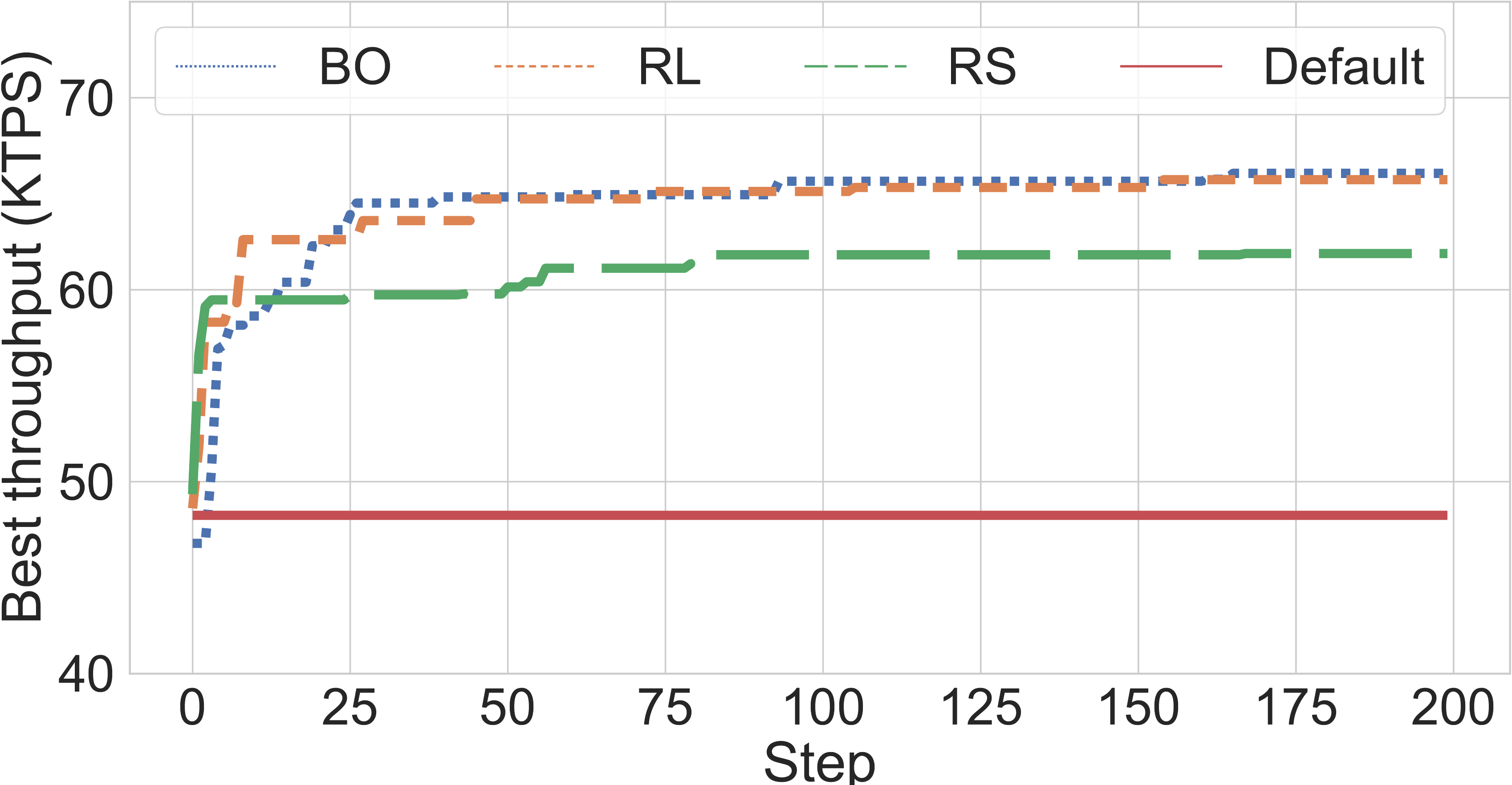}\caption{Throughput of the best configuration found over time (average of the three runs)}\label{fig1}
 \end{subfigure}
 \hfill
\begin{subfigure}[c]{0.49\textwidth}
\centering
 \includegraphics[scale=0.3]{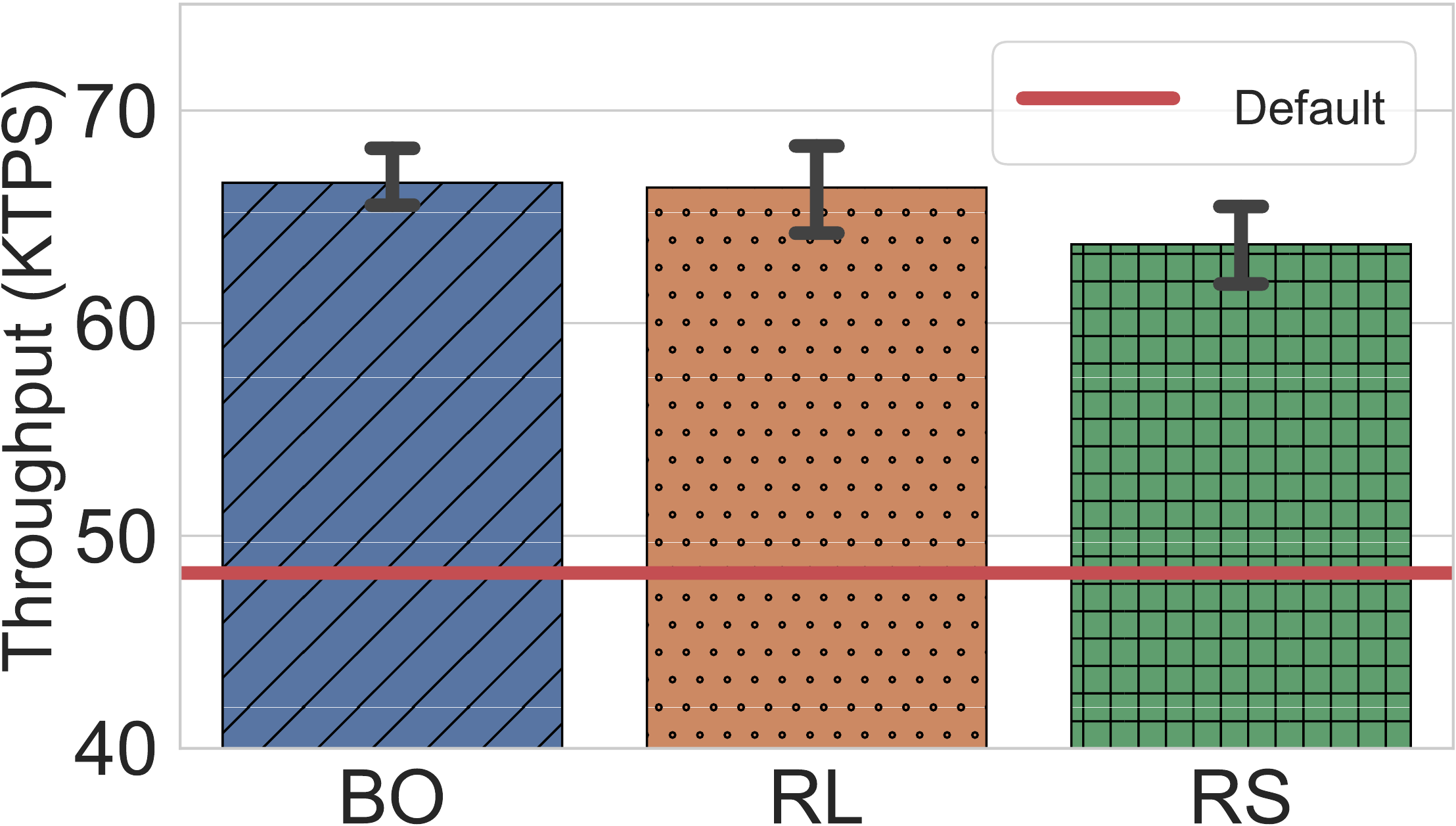}\caption{Average throughput and standard deviation corresponding to the best configurations identified by each optimization method.}\label{fig2}
\end{subfigure}

\caption{Experimental comparison of Bayesian optimization, reinforcement learning and random search}
\label{fig}
\end{figure*}

\noindent{\bf Results.} 
In Figure~\ref{fig1} we report the best throughput achieved (in thousands of transactions per second) as a function of the number of optimization steps, averaged over the three repetitions.
We observe that BO and RL have a similar behavior, finding a configuration with a higher throughput than random search.
Since evaluating the throughput of a given configuration is inherently noisy, we then took the best configuration found by each optimization run and re-evaluated it 5 times to obtain a more reliable estimate of the throughput, resulting in a total of 15 measurements for each optimization method. We report in Figure~\ref{fig2} the mean throughput and standard deviation achieved. We observe that, while BO and RL achieve a mean throughput of around 66 KTPs (38\% better than  the default configuration), simple random search achieves a mean throughput of 63.7 KTPs, which is only around 4\% lower.
In fact, when we apply the standard two-sided pairwise t-test we find that the null hypothesis (i.e., that there is no statistical difference between the configurations) cannot be rejected when comparing the configuration found by random search with that found by BO ($p>0.10$) and RL ($p>0.17$).

%% file: conclusion.tex

\section{Discussion}
We now describe the main insights we obtain from our study, and we outline some research directions.

~\\
\noindent{\bf i) The research community should focus on building self-tuning frameworks that work with as little knowledge as possible about the target database.} 
The main focus of existing research works is to identify the ML technique that achieves the best performance for a specific database taken as a use-case. 
We have shown, however, that there are challenges to the adoption of such techniques when applied to a new database. 
These challenges can jeopardize the optimization process altogether, making the effectiveness of the employed ML model a second-order concern.
Crucially, overcoming these challenges requires knowledge about the target database. We advocate that a primary goal of researchers should be making the adoption of self-tuning techniques easily applicable to new databases, without requiring domain knowledge, with the ultimate goal of delivering push-button optimization to the ordinary user.
Achieving this goal requires solving several issues, including automatically identifying the tuning parameters, determining their domains, and inferring valid value ranges, or designing optimization algorithms that are robust in the face of unknown ranges \cite{perrone2019learning, ha2019bayesian}.


~\\
\noindent{\bf ii) Databases should be built with self-tuning in mind.} Databases are extremely complex systems, and hence they are very hard to tune as a black box.  We advocate that databases should be built from the ground up with the goal of being amenable to self-tuning.  For example, database designers could annotate the tuning parameters to expose them and could define valid value ranges. The designers should also encode invariants that have to be maintained among parameters (e.g., one parameter has to be lower than another one) and specify when tuning a certain parameter has implications not only on performance but also on correctness (e.g., consistency and fault tolerance tuning parameters).

~\\
\noindent{\bf  iii) Model-free approaches warrant more attention.}
Our results indicate that random search is a competitive baseline in the context of self-tuning databases, achieving results comparable to more complex ML techniques such as BO and RL.
Similar results have been shown in recent work on hyper-parameter tuning~\cite{li2020random}, and random search is in general considered as a standard baseline by the ML community~\cite{bergstra2012random}.
However, random search is often not considered in the evaluation of  frameworks that use ML to optimize databases~\cite{ottertune,cdbtune}.
We argue that random search should always be considered as a baseline, since it helps to put performance gains into perspective.
Random search is also relatively easy to use since it is \textit{model-free}: it makes no assumptions regarding the structure of the unknown function being optimized.
In contrast, ML approaches such as BO and RL both rely on some mathematical model (typically Gaussian processes or neural networks respectively) leading to hyper-parameters that must be tuned.
Our results also suggest that other model-free optimization algorithms, such as successive halving~\cite{sa} and Hyperband~\cite{hb}, may be of significant interest in the context of self-tuning databases.


\section{Conclusions}
In this paper we detailed our experience in applying state-of-the-art ML-based self-tuning techniques to FoundationDB, a database with a plethora of configuration parameters not yet studied in this context.
We have identified critical issues faced when trying to apply these methods out-of-the-box to a new context, issues typically overlooked by prior work, and proposed ways to mitigate them; ways that should be applicable more-or-less unchanged to different database optimization targets.
We have presented experimental results when tuning FoundationDB using the two BO and RL based methods and compared them to a random search baseline.
Our results show that BO and RL methods can indeed improve the performance of FoundationDB by up to 38\% compared to the default configuration, but also show that, crucially, random search performs 4\% within the ML methods.
This echoes recent findings in neural architecture search, that show that random search is competitive with more complex approaches, suggesting that other model-free optimization methods should be investigated also in the context of searching the configuration space of databases.
